\documentclass[aps,superscriptaddress,twocolumn,showpacs,preprintnumbers,amsmath,amssymb]{revtex4}

\usepackage{mathrsfs}
\usepackage{amsmath}
\usepackage{amssymb}
\usepackage{graphicx}
\usepackage{dcolumn}
\usepackage{bm}
\usepackage{subfigure}
\usepackage{color}
\usepackage{ifpdf}
\usepackage{epstopdf}
\usepackage{CJK}
\usepackage{upgreek}
\usepackage[
  pdfstartview=FitH,%
  bookmarks=true,%
  bookmarksopen=true,%
  breaklinks=true,%
  colorlinks=true,%
  linkcolor=blue,anchorcolor=blue,%
  citecolor=blue,filecolor=blue,%
  menucolor=blue,pagecolor=blue,%
  urlcolor=blue]{hyperref}
\usepackage{float}

\begin{document}

\title{Period-doubled Bloch states in a Bose-Einstein condensate}
\author{Baoguo Yang}
\affiliation{School of Electronics Engineering and Computer Science, Peking University, Beijing 100871}
\author{Pengju Tang}
\affiliation{School of Electronics Engineering and Computer Science, Peking University, Beijing 100871}
\author{Xinxin Guo}
\affiliation{School of Electronics Engineering and Computer Science, Peking University, Beijing 100871}
\author{Xuzong Chen}
\affiliation{School of Electronics Engineering and Computer Science, Peking University, Beijing 100871}
\author{Biao Wu}\email{wubiao@pku.edu.cn}
\affiliation{International Center for Quantum Materials, School of Physics, Peking University, Beijing 100871}
\author{Xiaoji Zhou}\email{xjzhou@pku.edu.cn}
\affiliation{School of Electronics Engineering and Computer Science, Peking University, Beijing 100871}
\date{\today}

\begin{abstract}
We study systematically the period-doubled Bloch states for a weakly interacting Bose-Einstein condensate
in a one-dimensional optical lattice. This kind of state is of form $\psi_k=e^{ikx}\phi_k(x)$, where
$\phi_k(x)$ is of period twice the optical lattice constant.  Our numerical results show how these
nonlinear period-doubled states grow out of linear period-doubled states at a quarter away from the Brillouin zone center as the repulsive interatomic interaction increases. This is corroborated by our analytical results. We find that all nonlinear period-doubled Bloch states have both Landau instability and dynamical instability.
\end{abstract}

\maketitle

A Bose-Einstein condensate (BEC) in an optical lattice (OL) has been explored theoretically and experimentally for its rich physics~\cite{RevModPhys.80.885}, such as quantum phase transition~\cite{SuperMott,QT_Chen,QT_Lv,QT_Zhu}, unconventional superfluidity~\cite{WuCPB}, and various nonlinear effects~\cite{LiangCPL,WangCPL,DuanCPL,PRA66_013604,PRA67,PRA61_Wu,PRA65_025601,njp5_Wu,ChenZhu,PRA66_063603,PRA72_Holland, WFJCPL,PRL92,PRA80}.
In the mean-field theory, a BEC in an OL becomes a nonlinear periodic system, which exhibits many
features that can not be found in a linear periodic system. For example, the nonlinear periodic system can have loop structures in its Bloch band~\cite{PRA66_013604,PRA67,PRA61_Wu,PRA65_025601,njp5_Wu,ChenZhu,PRA66_063603,PRA72_Holland}.
Such a nonlinear system can have a type of solution called gap solitons, which are localized in space and whose chemical potential lies in the linear band gap~\cite{WFJCPL,PRL92,PRA80}.  These gap solitons can never exist in a linear periodic system. There is another type of solutions, which are Bloch-like states
but their periodic parts have a period that is twice the lattice constant~\cite{PRA69}.
These period-doubled states are closely related to the period-doubling phenomenon that has been
observed experimentally~\cite{PRL95_Chu}.

\begin{figure}[t]
  \includegraphics[width=0.48\textwidth]{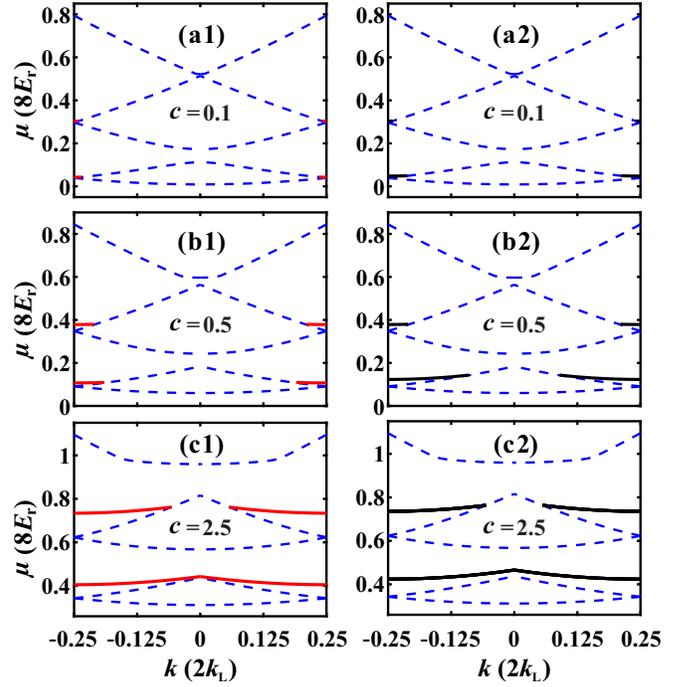}\\
  \caption{(color online) Energy bands for period-doubled Bloch states: type I on the left (red solid curves)
  and type II on the right (black solid curves).
  Energy bands for single-period  (blue dashed curves) are plotted for comparison. (a1,a2) $c=0.1$; (b1,b2)
  $c=0.5$; (c1,c2) $c=2.5$.  $V_0=1E_{\rm r}$.
  }\label{fig:Band_SD}
\end{figure}

In this Letter we study these period-doubled states systematically for a BEC in a one-dimensional OL.
We find that these states can be Bloch-like and the corresponding chemical potentials form Bloch-like bands.
Our numerical results show that when the repulsive interatomic interaction increases from zero
the period-doubled band begins to emerge around the quasi-momentum that is a quarter away from
the Brillouin zone center (see Fig.~\ref{fig:Band_SD}). As the interaction further increases, the period-doubled band
extends to the whole Brillouin zone. We have also analyzed the situation when the interaction is very small;
our analytical results are very consistent with our numerical results.
By computing their Bogoliubov spectrums, we further find all nonlinear period-doubled states  have Landau instability and dynamical instability.

We consider a weakly-interacting BEC in a one-dimensional OL. In the mean-field regime, this system
can be well described by the following Gross-Pitaevskii equation (GPE)~\cite{Gross-Pitaevskii}
\begin{eqnarray}\label{eqn:originalgpe3D}
i\hbar \dfrac{\partial \Phi\left( \mathbf{\tilde r},\tilde t\right) }{\partial \tilde t} &=&-\dfrac{\hbar ^{2}}{2m}{\nabla ^2}\Phi \left( \mathbf{\tilde r},\tilde t\right)+{\tilde V}(\mathbf{\tilde r})\Phi\left( \mathbf{\tilde r},\tilde t \right)   \nonumber \\
&&+\dfrac{4\pi \hbar ^{2}a_{\rm s}}{m}|\Phi\left( \mathbf{\tilde r},\tilde t\right) |^{2}\Phi\left( \mathbf{\tilde r},\tilde t\right),
\end{eqnarray}
where $a_{\rm s}$ is the s-wave scattering length, and $m$ is the atomic mass. We consider a cigar-shaped condensate~\cite{PRL87_130402,PRA65_063612}, so we can focus only on the lattice direction
and ignore the other directions. Mathematically, this is to write the matter wave function as
\begin{equation}\label{eqn:3Dwavefuncion_1D}
\Phi \left( {\mathbf{\tilde r},\tilde t} \right) = \varphi \left( {\tilde x,\tilde t} \right)\varphi _0 \left( \tilde y,\tilde z \right){e^{ - i\left( {{E_{\mathrm y}} + {E_{\mathrm z}}} \right)\tilde t/\hbar }},
\end{equation}
where $\varphi_0$ is the wave function in the transverse direction of the BEC with the corresponding energy $E_{\mathrm y} + E_{\mathrm z}$, which can be approximated by gaussian function. Thus, Eq.~(\ref{eqn:originalgpe3D}) can be reduced to a one-dimensional form
\begin{eqnarray}\label{eqn:originalgpe1D}
i\hbar \dfrac{\partial \varphi\left( \tilde x,\tilde t \right) }{\partial \tilde t}%
&=&-\dfrac{\hbar ^{2}}{2m}\dfrac{\partial^2 \varphi \left( \tilde x,\tilde t\right)}{\partial {\tilde x}^2}+V(\tilde x)\varphi\left( \tilde x,\tilde t\right)   \nonumber \\
&&+\dfrac{4\pi \hbar ^{2}a_{\rm s}}{mA}|\varphi\left( \tilde x,\tilde t\right) |^{2}\varphi\left( \tilde x,\tilde t\right),
\end{eqnarray}
where $A = 2/ \left| \varphi _0\left( 0,0 \right) \right|^2$ is the effective cross sectional area of the condensate. In the longitudinal direction, we do not consider the harmonic trap, and the OL potential is given by
\begin{eqnarray}\label{eqn:pot}
  V(\tilde x) = V_0 \cos^2(k_{\rm L} \tilde x) = \dfrac{V_0}{2} \cos(2 k_{\rm L} \tilde x) + \dfrac{V_0}{2},
\end{eqnarray}
where $k_{\rm L}=2 \pi/ \lambda$ with $\lambda$ the wavelength of the laser, and $V_0$ is the lattice depth. After neglecting the constant potential $V_0/2$, the dimensionless GPE is
\begin{eqnarray}\label{eqn:gpeN}
  i\dfrac{\partial \Psi \left(x,t\right)}{\partial t}&=&-\dfrac{1}{2}\dfrac{\partial^2 \Psi \left(x,t\right)}{\partial x^2} + \dfrac{v}{2} \cos(x)\Psi \left(x,t\right)\nonumber\\
  &&+ \dfrac{c}{8} |\Psi \left(x,t\right)|^2 \Psi \left(x,t \right),
\end{eqnarray}
where $\Psi \left( x,t \right) = \sqrt{\pi/(N_0 k_{\rm L})}\varphi \left( {\tilde x,\tilde t} \right)$, with $N_0$ the total number of atoms, $x$ is in units of $1/2k_{\rm L}$, $t$ is in units of $m/4\hbar k_{\rm L}^2$, and the potential well depth $v$ is in units of $8E_{{\rm r}}$ with $E_{{\rm r}}=\hbar^2 k_{\rm L}^2/2 m$ the recoil energy. The interaction constant $c={8\pi a_{\rm s} n_0/k_{\rm L}^2}$ with $n_0= N_0 k_{\rm L}/(\pi A)$ the averaged density of BEC. By substituting $\Psi \left( x,t \right) =\psi \left( x \right){e^{-i\mu t}}$ into Eq.~(\ref{eqn:gpeN}), we get the time-independent GPE
\begin{eqnarray}\label{eqn:ssgpe}
  &-\dfrac{1}{2}\dfrac{{\rm d}^2 \psi \left( x \right)}{{\rm d}x^2} + \dfrac{v}{2} \cos(x)\psi \left( x \right) + \dfrac{c}{8} |\psi \left( x \right)|^2 \psi \left( x \right)\nonumber\\
  &= \mu \psi \left( x \right),
\end{eqnarray}
where $\mu$ is the nonlinear eigenvalue. $\psi \left( x \right)$ satisfies the following normalization condition
\begin{equation}\label{eqn:psi_nor}
\int_{-l\pi}^{l\pi} {|\psi \left( x \right)|^2\,{\rm d} x}=2l\pi,
\end{equation}
with $l=1$ for single-period solutions and $l=2$ for period-doubled solutions.

\begin{figure}[t]
  \includegraphics[width=0.4\textwidth]{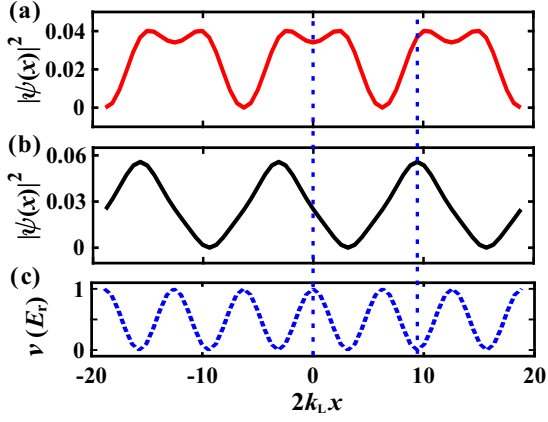}\\
  \caption{(color online)  Two types of  period-doubled Bloch states.  (a) Type I, whose peaks are around
  the crests of the lattice potential; (b) type II, whose peaks are around the troughs. The lattice potential
  is shown in (c).   Quasi-momentum $k=0.25$; OL depth $V_0=1E_{\rm r}$;  $c=0.5$.
  }\label{fig:Wave}
\end{figure}

When $c=0$, the solutions of Eq.(\ref{eqn:ssgpe}) are usual Bloch waves, which can be
expressed as
\begin{equation}
\label{eq:lin}
\psi_k^0(x)=e^{ikx}u_k(x)=e^{ikx}u_k(x+2\pi),
\end{equation}
where $\hbar k$ is the quasi-momentum.  When $c\neq 0$, besides these
usual Bloch waves, there exist solutions of the following form
\begin{eqnarray}\label{eqn:blochwave}
  \psi_k(x)={\rm e}^{ikx}\phi_k(x)={\rm e}^{ikx}\phi_k(x+4\pi).
\end{eqnarray}
We call these solutions period-doubled Bloch waves.
By substituting Eq.~(\ref{eqn:blochwave}) into Eq.~(\ref{eqn:ssgpe}), we have
\begin{eqnarray}\label{eqn:phieqn}
  &-\dfrac{1}{2}\left(\dfrac{{\rm d}}{{\rm d}x}+ik\right)^2 \phi_k + \dfrac{v}{2} \cos(x)\phi_k + \dfrac{c}{8} |\phi_k|^2 \phi_k\nonumber\\
  &= \mu(k) \phi_k.
\end{eqnarray}
These period-doubled Bloch states $\phi_k(x)$ can be
expanded in the following Fourier series
\begin{equation}\label{eqn:fourierexpansion}
  \phi_k(x) = \sum_{n=-N}^{N}a_n {\rm e}^{i \frac{n}{2} x},
\end{equation}
where $N$ is the cut-off. By substituting Eq.~(\ref{eqn:fourierexpansion}) into Eq.~(\ref{eqn:phieqn}), we have
\begin{eqnarray}\label{eqn:an}
  &\dfrac{1}{2}\left(\dfrac{n}{2}+k\right)^2 a_n + \dfrac{v}{4}\left(a_{n-2}+a_{n+2}\right)\nonumber\\
  &+\dfrac{c}{8} \displaystyle\sum_{n_1=-N}^{N}\displaystyle\sum_{n_2=-N}^{N}a_{n_1}^* a_{n_2} a_{n -n_2 +n_1} = \mu a_n.
\end{eqnarray}

Numerically solving the above equations for $a_n$ and $\mu$ together with the normalization
condition $\sum_{n=-N}^{N} |a_n|^2 = 1$, we can find both the single-period and period-doubled solutions.
The results are plotted in Fig.~\ref{fig:Wave} and Fig.~\ref{fig:Band_SD} with OL depth $V_0=1 E_{\rm r}$.
For the period-doubled Bloch waves, their Brillouin zone is only half of the
Brillouin zone of the usual single-period Bloch waves. Therefore,
to compare between the usual Bloch states and period-doubled solutions, we have folded up
the first Brillouin zone for the usual Bloch states in Fig.~\ref{fig:Band_SD}.

There are two types of period-doubled Bloch states as shown in Fig.~\ref{fig:Wave}.
For type I states, mathematically their coefficients $a_n$'s are all real; physically,
their peaks are around the crests of the lattice potential (see Fig.~\ref{fig:Wave}(a)).
For type II,  mathematically their coefficients $a_n$'s are real for even $n$ and
pure imaginary for odd $n$; physically,  their peaks are around the troughs of the
lattice potential (see Fig.~\ref{fig:Wave}(b)).

\begin{figure}
  \includegraphics[width=0.48\textwidth]{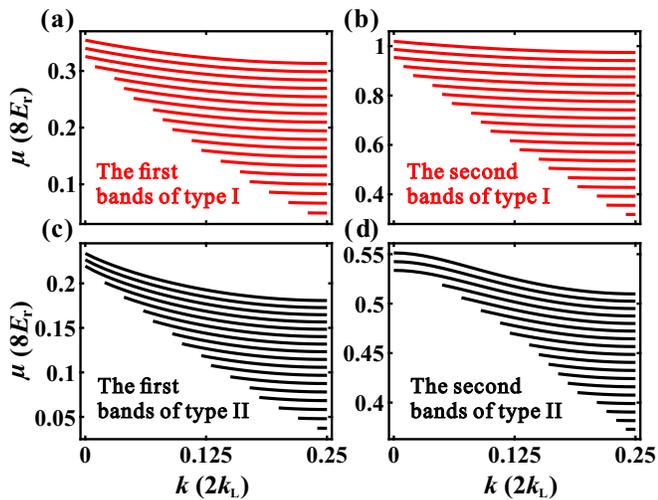}\\
  \caption{(color online) Period-doubled energy bands for different atomic interaction $c$.
  The upper two panels (a,b) are for type I period-doubled Bloch states; the lower two panels (c,d) are for type II states.
  (a) From bottom to top, the first bands of type I
  respectively for $c$ ranging from $0.1$ to $1.8$ with a step size of $0.1$; $V_0=0.1E_{\rm r}$.
  (b) From bottom to top, the second bands of type I  respectively for $c$ ranging from $0.2$ to $4$
  with a step size of $0.2$; $V_0=0.1E_{\rm r}$.
  (c) From bottom to top, the first bands of type I  respectively for $c$ ranging from $0.05$ to $0.85$ with a step size of $0.05$;
  $V_0=1E_{\rm r}$.
  (d) From bottom to top, the second bands of type II
  respectively for $c$ ranging from $0.04$ to $0.55$ with a step size of $0.03$; $V_0=10E_{\rm r}$. }
  \label{fig:Band_D}
\end{figure}

Similar to the usual Bloch waves, the period-doubled Bloch states can also form energy bands. These energy bands in terms of
chemical potential $\mu(k)$ are plotted in Fig.~\ref{fig:Band_SD}, where they are compared to the usual Bloch bands.
It is clear from Fig.~\ref{fig:Band_SD} that the period-doubled Bloch bands start at $k=1/4$, where the fold-up single-period Bloch
bands are degenerate. As the interaction gets stronger, that is, $c$ increases,  the period-doubled bands extend further toward
the Brillouin zone center. Around $c=2.5$, the lowest period-doubled bands for both type I and type II are extended over the full
Brillouin zone.  We have plotted these period-doubled bands for different $c$ together in  Fig.~\ref{fig:Band_D}, where
one can see more clearly how the bands grow as $c$ increases.  Or, one can view it reversely and see how these
period-doubled Bloch bands disappear when the interaction $c$ is reduced to zero. These numerical results strongly
suggest that   the period-doubled Bloch states are grown out of some ``seeds" in the usual single-period Bloch states.
In the following we take a closer look analytically.

Consider the linear case $c=0$ and the lowest band. With the usual Bloch waves, we can construct period-doubled states as follows
\begin{eqnarray}
\varphi^0_k(x)&=&c_1\psi_k^0(x)+c_2\psi_{k-\frac12}^0(x)\nonumber\\
&=&e^{ikx}\big[c_1u_{k}(x)+c_2e^{-ix/2}u_{k-\frac12}(x)\big]\,,
\label{eq:lind}
\end{eqnarray}
where $c_1$ and $c_2$ are complex and satisfy $|c_1|^2+|c_2|^2=1$. These period-doubled states $\varphi^0_k(x)$
are solutions of Eq.(\ref{eqn:ssgpe}) with $c=0$ only when $k=1/4$. Our numerical results show that
as $c$ decreases to zero the nonlinear period-doubled Bloch states in the lowest bands shown in Fig.~\ref{fig:Wave}, ~\ref{fig:Band_SD} \&~\ref{fig:Band_D} are reduced to $\varphi^0_{\frac14}(x)$ with specified $c_1$ and $c_2$.  To specify $c_1$ and $c_2$, we need to
fix the phases of the usual Bloch states $\phi^0_k(x)$. In Eq.~\ref{eqn:fourierexpansion}, our phase convention is taken in such a way
that the largest coefficient $a_m$ is real and positive. With this phase convention,
according to our numerical calculation, type I is connected to
\begin{equation}
\label{eq:I}
\varphi^0_{1/4}(x) = \frac{1}{{\sqrt 2 }}\left[ {\psi _{1/4}^{0}( x ) \pm \psi _{-1/4}^{0}( x )} \right]
\end{equation}
 and type II is to
 \begin{equation}
 \label{eq:II}
 \varphi^0_{1/4}( x ) = \frac{1}{{\sqrt 2 }}\left[ {\psi _{1/4}^{0}( x ) \pm i\psi _{-1/4}^{0}( x )} \right].
 \end{equation}
The results are similar for the second or higher bands of period-doubled Bloch states.

The above numerical results have given us clear guidance on how to obtain some analytical results, in particular, when $c$ is small.
When $c$ increases slightly from zero, we observe that two things will happen. (i) The period-doubled states  $\varphi^0_{1/4}( x )$ will persist with slightly
modified form. (ii) New period-doubled states slightly away from $k=1/4$ will emerge. When $k\neq 1/4$ and $c=0$, the states
$\varphi^0_{k}( x )$ in Eq.(\ref{eq:lind}) are not solutions of Eq.(\ref{eqn:ssgpe}) due to that $\psi^0_{k}$ and $\psi^0_{k-\frac12}$
have different eigen-energies. When $c$ is not zero, the interaction energy may bridge this energy gap and render $\varphi^0_{k}( x )$
be the solutions of Eq.(\ref{eqn:ssgpe}).

\begin{figure}
  \includegraphics[width=0.48\textwidth]{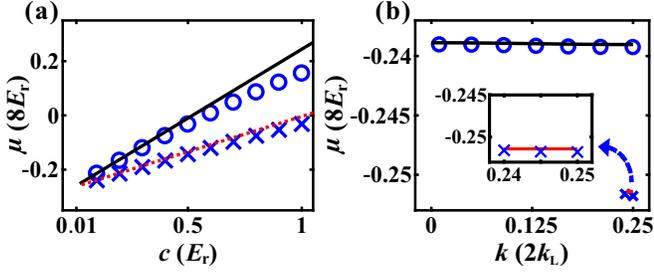}\\
  \caption{(color online) Comparison between analytical results and numerical results for
  chemical potentials of period-doubled Bloch states. The analytical results are given by red dashed curves
  for type I and black solid curves for type II; and the numerical ones are given by blue circles and crosses,
  respectively.  (a) Chemical potentials at $k=1/4$ as functions of interaction strength $c$.
  (b) Period-doubled energy bands for $c=0.05$.  As the band for type I is very narrow, it is zoomed up in the inset.
  $V_0=10E_{\rm r}$.}\label{fig:Analysis}
\end{figure}

Based on observation (i), we expect the nonlinear period-doubled Bloch state to have the following form
\begin{eqnarray}\label{eqn:Appr1}
 {\psi _{\frac14}}\left( x \right) &\approx& \sqrt {\frac{1}{2}{\rm{ - }}\delta^2} \left[  {{\psi^{0}_{1,\frac14}}\left( x \right) \pm e^{i{\theta_0}}
 {\psi^{0}_{1,-\frac14}}\left( x \right)} \right] \nonumber \\
 && + {\delta}\left[ {{\psi^{0}_{2,\frac14}}\left( x \right) \pm e^{i{\theta_0}}  {\psi^{0}_{2,-\frac14}}\left( x \right)} \right],
\end{eqnarray}
where $\theta_0 = 0$ for type I and $\theta_0 = \pi/2$ for type II. The integers in the subscript of $\psi^{0}$ are band indices
as the states in the second linear Bloch band are involved due to interaction. $\delta$ is small and has the same order of magnitude of $c$.
By substituting the above $\psi_{1/4}(x)$ into Eq.~\ref{eqn:ssgpe}, and keeping to the first-order correction, we obtain
\begin{eqnarray}\label{eqn:mu1}
 \mu_{\frac14}  = {\mu^{0}_{1,\frac14}} + \frac{c}{{64\pi }}\int_{ - \pi }^\pi  {{{\left| {{\psi^{0}_{1,\frac14}} \pm e^{i{\theta_0}}{\psi^{0}_{1, -\frac14}}} \right|}^4}{\rm{d}}x}.
\end{eqnarray}
This result is plotted in Fig.~\ref{fig:Analysis}(a) as a function of $c$ with
red dashed curve for type I and black solid curve for type II. The corresponding numerical results are shown with blue crosses and circles, respectively. It is clear from Fig.~\ref{fig:Analysis}(a) that our approximation above is reasonably good.

For observation (ii), we consider period-doubled states with $k$ less than but close to $1/4$.
In this case we only need to consider the lowest linear band and approximate the nonlinear period-doubled states as
\begin{eqnarray}\label{eqn:Appr2}
 {\psi _{k}}\left( x \right) = \sqrt {\frac{1}{2} + {\delta}} { \psi^{0}_{k}}\left( x \right) \pm e^{i{\theta_0}}\sqrt {\frac{1}{2} - {\delta}} {\psi^{0}_{k-\frac12}}\left( x \right).
\end{eqnarray}
By Substituting Eq.(\ref{eqn:Appr2}) into Eq.(\ref{eqn:ssgpe}), we find that the
chemical potential can be approximated as
\begin{eqnarray}\label{eqn:mu2}
 \mu_{k} &=& \frac{\mu^{0}_{k} + \mu^{0}_{k-\frac12}}{2} \nonumber \\
 &&+ \frac{c}{64\pi }\int_{ - \pi }^\pi  \left|  {\psi^{0}_{k}} \pm e^{i{\theta_0}} {\psi^{0}_{k-\frac12}}
 \right|^4{\rm{d}}x.
\end{eqnarray}
This analytical result is compared to the corresponding numerical ones in Fig.~\ref{fig:Analysis}(b),
where the red curve is for $\theta_0 = 0$ and the black one is for $\theta_0 = \pi/2$.
The corresponding numerical results are marked with crosses and circles. We see that
they are very consistent with each other.

The stability of the usual Bloch states has been studied both theoretically and experimentally~\cite{PRA67,PRA64,njp5_71,PRL93_Inguscio,PRA93_Inguscio,PRL86_1402,PRE63,PRL89_170402}.
It was found that many of the usual Bloch states near the Brillouin zone edge $\pm k_{\rm L}$ are unstable, suffering both Landau instability and dynamical instability. It is worthwhile and also necessary
to examine the stability of these period-doubled Bloch states.

As the details of how to examine the Landau instability and dynamical instability has been spelled out
in literature~\cite{njp5_Wu}, we just briefly summarize the procedure.  To study Landau instability,
we need to compute the eigenvalues of the following matrix
\begin{eqnarray}\label{eqn:mkq_x}
  M_k(q) = \left(
  \begin{array}{ccc}
  \mathscr{L}(k+q) & \dfrac{c}{8}\phi_k^2 \\
  \dfrac{c}{8}{\phi_k^*}^2 & \mathscr{L}(-k+q)
  \end{array} \right)
\end{eqnarray}
where $q$ is the perturbation Bloch wave number, and
\begin{eqnarray}\label{eqn:lk}
  \mathscr{L}(k) &=& -\dfrac{1}{2}\left(\dfrac{\partial}{\partial x}+ik\right)^2+\dfrac{v}{2}\cos(x)\nonumber\\
  &&+\dfrac{c}{4}|\phi_k(x)|^2-\mu.
\end{eqnarray}
Diagonalizing the matrix $M_k(q)$, we can get the eigenvalues. If $M_k(q)$ is positive definite for all $-0.25 \leq q \leq 0.25$ for period-doubled solutions, the solution $\phi_k(x)$ is a local minimum and has no Landau instability. If $M_k(q)$ has negative eigenvalues for some $q$, the Bloch wave is a saddle point and suffers Landau instability.

For dynamical instability, we need to diagonalize another matrix ${\sigma_z}{M_k}(q)$, where
\begin{eqnarray}\label{eqn:sz}
{\sigma _z} = \left( {\begin{array}{*{20}{c}}
I&0\\
0&{ - I}
\end{array}} \right).
\end{eqnarray}
If all eigenvalues of $\sigma_z M_k(q)$ are real for all $-0.25 \leq q \leq 0.25$ for period-doubled solutions,
the period-doubled state is dynamically stable. If there are complex eigenvalues, the initial small disturbance can grow exponentially in time, the state is dynamically unstable. We denote the maximum values among the imaginary
part of eigenvalues of matrix $\sigma_z M_k({q})$ as $M_{\rm D}$.

\begin{figure}
  \includegraphics[width=0.48\textwidth]{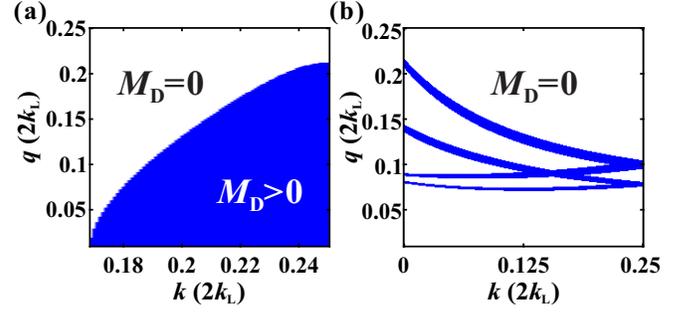}\\
  \caption{(color online) Dynamical stability phase diagrams for (a) period-doubled Bloch states of type I in the first band
   with $c=0.4$; (b) period-doubled states of type I of the second band with $c=4$.   $M_{\rm D}$ denotes the maximum values among the imaginary part of eigenvalues of Matrix $\sigma_z M_k({q})$; the blank ares are for
   $M_{\rm D}=0$. $V_0=0.1E_{{\rm r}}$.}\label{fig:Insta}
\end{figure}

In our calculation we consider the states in Fig.~\ref{fig:Band_D}.
We find that all period-doubled states of type II (shown in Fig.~\ref{fig:Band_D}(c) and Fig.~\ref{fig:Band_D}(d)) have both Landau instability and dynamical instability. All the period-doubled Bloch states of type I shown in both Fig.~\ref{fig:Band_D}(a) and Fig.~\ref{fig:Band_D}(b) have Landau instability.
The type I  states in the first band with higher nonlinear interaction $c$ in Fig.~\ref{fig:Band_D}(a) have dynamical instability.  However, when $c$ is small, for a given $k$ there are always values of $q$
where $M_{\rm D}=0$, corresponding to the blank areas shown in Fig.~\ref{fig:Insta}(a). There are also cases of $M_{\rm D}=0$ for type I states
in the second bands  for different $c$ as  shown in Fig.~\ref{fig:Insta}(b).
For the reflection symmetry of results in $k$ and $q$, only the parameter region of $0 \leq k \leq 1/4$ and $0 \leq q \leq 1/4$ is shown.  Nevertheless, as it is hard to control the modes of perturbations in actual experiments, from the point of experiments, these period-doubled states have dynamical instability.

In conclusion, we have systematically studied period-doubled Bloch states for  a BEC in a one-dimensional
optical lattice. Both our numerical and analytical results show that these period-doubled Bloch states
can be viewed as growing out of ``seeds" which are linear period-doubled states. We have found
that all these period-doubled Bloch states suffer both Landau instability and dynamical instability.
Some phenomena related period-doubled Bloch states have been observed experimentally~\cite{PRL95_Chu}.
However, it is still quite challenging to observe these states  directly and clearly in a controlled way due to
that these period-doubled Bloch states are not stable. It would be interesting in the future to find stable period-doubled
Bloch states by engineering the nonlinear interaction.

This work is supported by the National Key Research and Development Program of China (Grant Nos. 2016YFA0301501, 2017YFA0303302),
and the National Natural Science Foundation of China (Grants Nos.11334001, 61727819, 61475007, 91736208).

\end{document}